\documentclass[
 amsmath,amssymb,
 aps, physrev,
]{revtex4-2}

\usepackage{graphicx}
\usepackage{dcolumn}
\usepackage{bm}
\usepackage{epstopdf}
\begin{document}

\preprint{APS/123-QED}

\title{\textbf{Space Charge-Induced Emittance Growth in the downstream Section of ERL Injectors.} }%

\author{Xiuji Chen}
\email{chenxj2@shanghaitech.edu.cn}
\affiliation{ShanghaiTech University, Shanghai 201210, China}
\affiliation{Shanghai Institute of Applied Physics, Chinese Academy of Sciences, Shanghai 201800, China}

\author{Zipeng Liu}
\email{Contact author: liuzp@zjlab.ac.cn}
\affiliation{Zhangjiang Laboratory, Shanghai 201200, China}

\author{Si Chen}
\affiliation{Shanghai Advanced Research Institute, Chinese Academy of Sciences, Shanghai 201210, China}

\author{Duan Gu}
\affiliation{Shanghai Advanced Research Institute, Chinese Academy of Sciences, Shanghai 201210, China}

\author{Houjun Qian}
\affiliation{Zhangjiang Laboratory, Shanghai 201200, China}

\author{Dong Wang}
\affiliation{Shanghai Advanced Research Institute, Chinese Academy of Sciences, Shanghai 201210, China}

\author{Haixiao Deng}
\affiliation{Shanghai Advanced Research Institute, Chinese Academy of Sciences, Shanghai 201210, China}

\date{\today}

\begin{abstract}
The injector for ERL-FEL has been widely researched. Unlike traditional linacs, the bunch in the injector for ERLs requires additional deflection and matching section at lower energies. It makes the bunch more susceptible to the effects of the Space Charge. This will lead to a degradation in beam quality. In this paper, we comprehensively analyze the impact of space charge on ERL-injector and propose new design concepts to further maintain beam quality.

\end{abstract}

\maketitle


\section{introduction}
Energy Recovery Linac (ERL) injectors play a crucial role in the operation of ERLs, serving as the source and initial acceleration stage for electron beams. An ERL injector can be divided into three main sections: the injection section, the matching section, and the merger section\cite{angal2018perle,bilderback2003energy,klein2022energy}. ERL injectors often operate in the space charge regime to enhance energy recovery efficiency. Consequently, the electron bunches are significantly influenced by space charge forces. This influence results in emittance growth, consequently limiting the brightness of the electron bunches. The space charge effects can be divided into longitudinal and transverse directions. The space charge in ERL injectors can be categorized into two components: transverse space charge (TSC) and longitudinal space charge (LSC). The nonlinear TSC induces a nonlinear distribution in the transverse phase space, leading to variations in slice emittance\cite{Mizuno2015,MIZUNO2022166733}. The LSC does not directly cause transverse emittance growth. Instead, it primarily affects the distribution in the longitudinal phase space\cite{khan2018space,venturini2008models}. However, in beamlines with non-zero dispersion, such as mergers, the additional energy spread caused by LSC couples with the transfer matrix. This coupling leads to displacements in the transverse plane, consequently resulting in growth of the projected emittance\cite{kayran2005method,litvinenko2006merger,hwang2012effects}. 

For the injection section, the combination of Multi-Objective Genetic Algorithm (MOGA) with finite element methods is an effective design approach\cite{abo2012astra,todd2006state,bazarov2003high}. In most cases, the bunch parameters at the exit of the injection section are used as optimization objectives. The space charge effects in the subsequent sections, such as the matching section and merger section, are not considered in this design work. It will lead to emittance dilution in these sections. In this paper, two methods are proposed to reduce the effects caused by TSC and LSC in the downstream sections of ERL injectors. In Sec.\ref{S2}, a pre-modulation optimization method is proposed based on the studies in Ref\cite{Mizuno2015,MIZUNO2022166733}. We quantitatively estimated the pre-modulation range and validated this method through simulations. In Sec.\ref{S3}, a quantitative method to estimate LSC effects in the merger section (or the arc section in low-energy machines) is proposed. The previous merger design is further optimized based on this method.

\section{the pre-modulation optimization method for the TSC in the downstream sections of ERL injectors}\label{S2}

For the beamline where only axisymmetric electromagnetic fields are present, the transverse emittance can be replaced by\cite{MIZUNO2022166733}:

\begin{equation}
\label{EQ:emit}
\varepsilon_x=\sqrt{\left<x^2\right>\left<x'^2\right>-\left<xx'\right>^2}=\frac{1}{\sqrt{2}N_e}\sqrt{\sum_{i,j=1}^{N_e}\left(X_i^T\hat{S}X_{j}\right)^2}
\end{equation}
$\varepsilon_x$ is the emittance. $(x,x')$ is the any point in horizontal plane. Where $x'_i=P_x/P_s$ in this paper. $P_i$ is the momenta in $i$-axis. $X_i=(x_i, x'_i)^T$ is the coordinate of particle $i$ in the bunch. $N_e$ is the total numbers of the particles in the bunch. $\hat{S}$ is the standard symplectic matrix. Notice the $X^T_i\hat{S}X_j=\vec{x}_i\times\vec{x}_j$ is the area of the quadrilateral formed by $\vec{x}_i$ and $\vec{x}_j$. $\vec{x}_i$ is the vector form the original point to the particle $i$. And it is easy to prove the area for this quadrilateral is an invariant under symplectic transformations. The $k_i=x'_i/x_i$ is the slop of $\vec{x}_i$. When all particles are linearly aligned, $k_i = k_j$ for any $i$ and $j$, the area equals zero. And the correlation between $x$ and $x'$ approaches 1, the $\varepsilon_x$ equals 0.

\begin{figure} 
\begin{center}
\includegraphics[height=5cm]{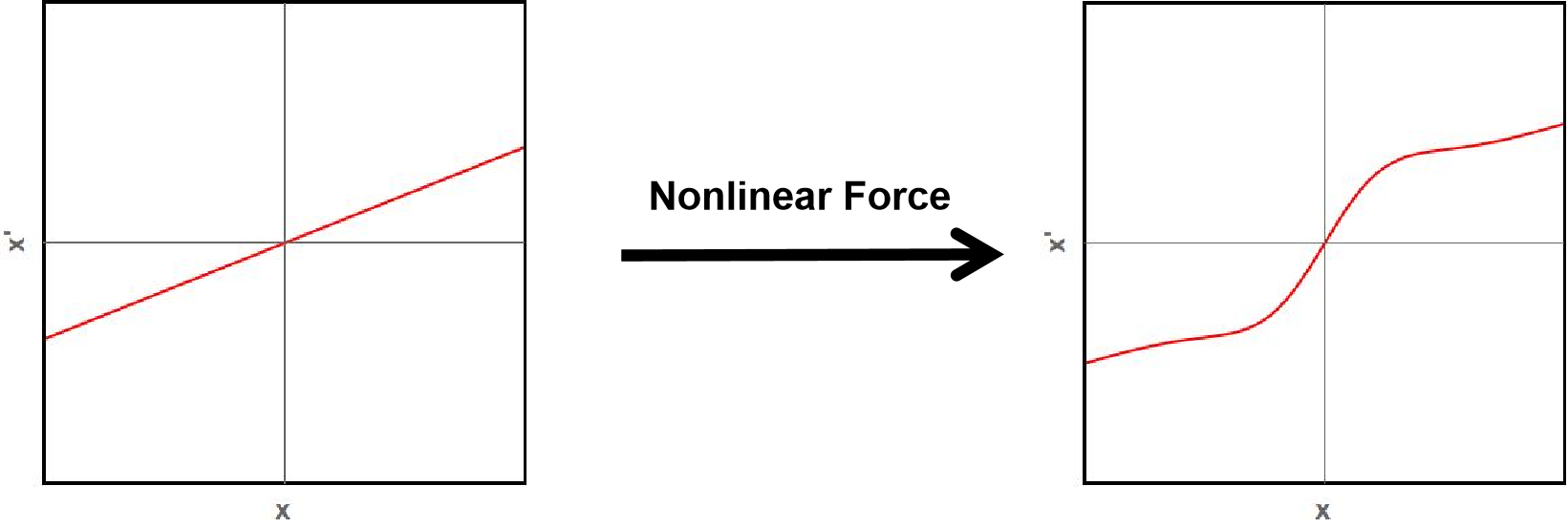} \\
\includegraphics[height=5cm]{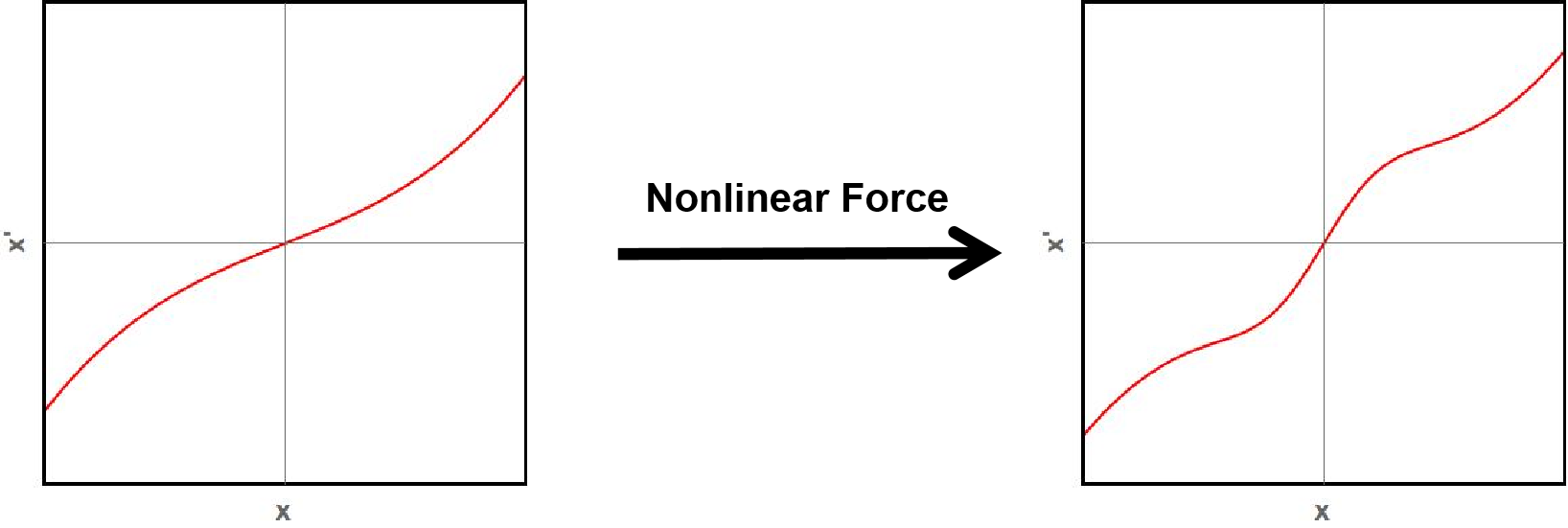}
\caption{\label{fig:model}The model for the growth in emittance caused by the nonlinear force (top) and the pre-modulation method to reduce the effects caused by the nonlinear force (bottom).}
\end{center}
\end{figure}

Based on the content discussed above, the model for the variance in emittance was shown in Fig.\ref{fig:model}(top): a nonlinear force acts on the particles, causing $k_i > k_j$ (or $k_i < k_j$), which leads to variance in $(X_i^T \hat{S} X_j)^2$. To reduce the effects caused by the nonlinear force, a pre-modulation method was proposed. The simple model is shown in Fig.\ref{fig:model} (bottom): a pre-modulation causes the bunch to meet the condition $k_i < k_j$ or $k_i > k_j$. This pre-modulation counteracts the nonlinear force, which reduces the variance in slice emittance. A more detailed description can be found in Ref.\cite{Mizuno2015}.

\begin{figure} 
\begin{center}
\includegraphics[height=6cm]{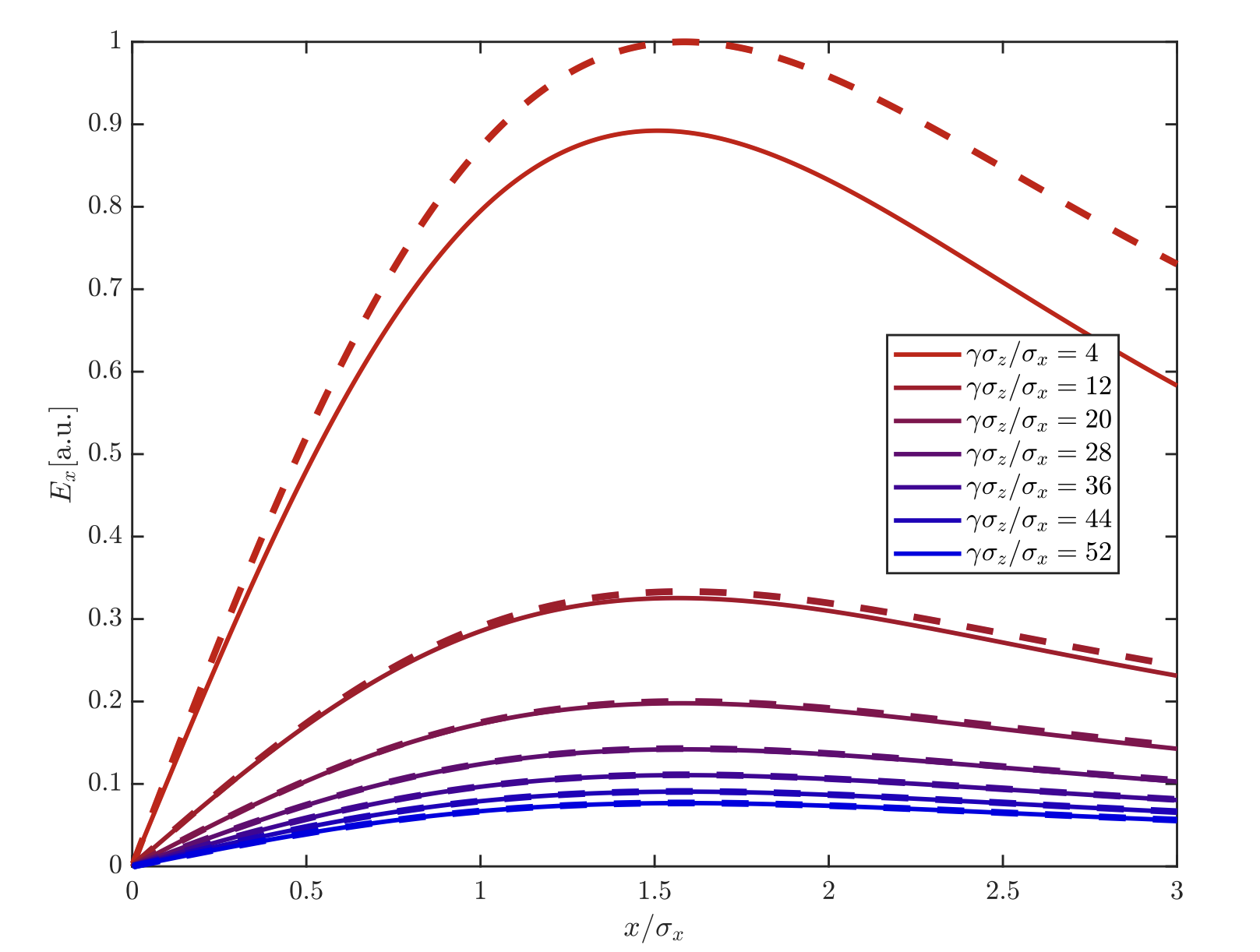} 
\caption{\label{fig:Er-anavsnum}
The approximate analysis of Eqs.\ref{EQ:approx_E} (dashed line) and the field calculated by numerical integration (solid line) where $y=0$, $z=0$.}
\end{center}
\end{figure}

For the ERL-injector, the nonlinear force originates from TSC effects. Furthermore, the mean emittance $<W_{ij}>$ was proposed to evaluate the degree of nonlinearity in phase space\cite{MIZUNO2022166733}:

\begin{equation}
\label{EQ:wij}
\left<W_{ij}\right>=\left<\vec{x_i}\times\vec{x_j}\cdot \frac{\vec{x_i}-\vec{x_j}}{|\vec{x_i}-\vec{x_j}|}\right>.
\end{equation}
The $\left<W_{ij}\right>$ was simplified as $W$ in this study. For the case where $k_i < k_j$, we have $W_{ij} < 0$; otherwise, the opposite is true. And the variance in mean emittance caused by any force can be expressed as
\begin{equation}
\label{EQ:detw}
\frac{d W_f}{ds}=\left<x_i\vec{f}(x_j)-x_j\vec{f}(x_i)\right>_{x_i,x_j>0}.
\end{equation}
Where $\vec{f}=\frac{\vec{F}}{\gamma m_e c^2}$ is the normalized force. $\gamma$ the relativistic Lorentz factor, $m_e$ the electron rest mass, and $c$ the speed of light. Next, we attempt to quantitatively analyze the variance in mean emittance caused by the TSC (transverse space charge) force in the downstream section of injectors. In the following derivation, we need to apply the Lorentz transformation:
\begin{equation}
\label{EQ:lor_trans}
\tilde{z}=\gamma z  \quad \tilde{\sigma}z=\gamma \sigma_z  \quad \tilde{E}_{x, y}=\gamma E_{x, y}  \quad \tilde{E}_z=E_z.
\end{equation}
In this equation, parameters with a tilde superscript represent the bunch parameters in the beam frame, while the others are parameters in the lab frame. For Gaussian bunches that satisfy the condition $\sigma_x=\sigma_y<<\gamma\sigma_z$, the TSC electric field can be expressed by the following formula by using the Gauss's law: 
\begin{equation}
\label{EQ:approx_E}
\tilde{E}_x(x, r, \tilde{z})=\frac{1}{2 \pi \epsilon_0} \frac{N_e ex}{\sqrt{2 \pi}r \tilde{\sigma}_z} e^{-\frac{\tilde{z}^2}{2 \tilde{\sigma}_z^2}}\left[\frac{1-e^{-\frac{r^2}{2 \sigma_x^2}}}{r}\right],
\end{equation}
where $r=\sqrt{x^2+y^2}$, $N_e$ is the number of electrons, $e$ is the elementary charge, $\epsilon_0$ is the vacuum permittivity, $\tilde{\sigma}_z$ is the RMS bunch length in beam frame, $\sigma_x$ is the RMS transverse beam size. Fig.\ref{fig:Er-anavsnum} shows the comparison between Eqs.\ref{EQ:approx_E} and the results of numerical integration. For a Gaussian bunch, the numerical integration can be simplified to\cite{takayama1982new}:

\begin{equation}
\label{EQ:numerical_int}
\tilde{E}_x = \frac{N_e e}{4 \pi \epsilon_0 \sqrt{\pi}} \int_0^{\infty} d q \frac{e^{-\frac{x^2}{2 \sigma_x^2+q} -\frac{y^2}{2 \sigma_y^2+q} -\frac{\tilde{z}^2}{2 \tilde{\sigma}_z^2+q}}}{\sqrt{\left(2 \sigma_x^2+q\right)\left(2 \sigma_y^2+q\right)\left(2 \tilde{\sigma}_z^2+q\right)}} \cdot \frac{2x}{2 \sigma_x^2+q}.
\end{equation}
As shown in Fig.\ref{fig:Er-anavsnum}, the approximate analysis agrees well with the numerical integration when $\gamma\sigma_z/\sigma_x > 10$. For ERL injectors, this is a reasonable operating range. And the normalized force for TSC effects is
\begin{equation}
\label{EQ:detmomenta}
\vec{f}_{x,sc}=\frac{e \vec{E}_x}{\gamma^3m_ec^2}
\end{equation}
Using Eqs.\ref{EQ:detw} and \ref{EQ:detmomenta}, we can derive the variance in mean emittance caused by TSC effects. Furthermore, the variance can be separated into two integrals:
\begin{equation}
\label{EQ:Iz}
I_z=\int_{-\infty}^{+\infty}\lambda(z)\lambda(\gamma z)dz=\frac{1}{2\sqrt{\pi}\gamma\sigma_z},
\end{equation}
\begin{equation}
\label{EQ:It}
\begin{split}
I_T = &\int^{+3\sigma_x}_{-3\sigma_x}\lambda(y_1)\int_{0}^{+3\sigma_x}\hat{\lambda}(x_1)\int_{x_1}^{+3\sigma_x}\left[\frac{1-e^{-\frac{r_1^2}{2 \sigma_x^2}}}{r_1}\right]\frac{x_1}{r_1}x_2\hat{\lambda}(x_2) dx_2dx_1dy_1 \\
&- \int^{+3\sigma_x}_{-3\sigma_x}\lambda(y_2)\int_{0}^{+3\sigma_x}\hat{\lambda}(x_1)\int_{x_1}^{+3\sigma_x}\left[\frac{1-e^{-\frac{r_2^2}{2 \sigma_x^2}}}{r_2}\right]\frac{x_2}{r_2}x_1\hat{\lambda}(x_2) dx_2dx_1dy_2
\end{split}
\end{equation}

\begin{equation}
\label{EQ:Iall}
\left<\frac{d W_{sc}}{ds}\right>_{x_i,x_j>0}=\mathcal{W}'_{sc}=\frac{N_er_e}{\gamma^2}I_zI_T.
\end{equation}
where $r_e$ is the classical electron radius, and
\begin{equation}
\label{EQ:lambda1}
\lambda(x_i)=\frac{1}{\sqrt{2\pi}\sigma_{x}}e^{-\frac{x_i^2}{\sigma_x^2}},
\end{equation}

\begin{equation}
\label{EQ:r}
r_i=\sqrt{x_i^2+y_i^2}.
\end{equation}
Notice only $x>0$ was considered in the mean emittance of the study, then the distribution in $x$ is
\begin{equation}
\label{EQ:lambda}
\hat{\lambda}(x)=\frac{\sqrt{2}}{\sqrt{\pi}\sigma_{x}}e^{-\frac{x^2}{\sigma_x^2}}.
\end{equation}
For the integral involving the longitudinal distribution, an analytical solution is straightforward. However, it is difficult to analytically solve the integral related to the transverse distribution. But it can be proved that this integral is independent of $\sigma_x$ and $\sigma_y$ for the round beam approximation (i.e., $\sigma_x = \sigma_y$): By substituting the integration variables $x_i$, $y_i$, $r_i$ with $u_{x_i}=\frac{x_i}{\sigma_{x}}$, $u_{y_i}=\frac{y_i}{\sigma_{x}}$ and $u_{r_i}=\frac{r_i}{\sigma_x}$ , the $I_T$ can be replaced by
\begin{equation}
\label{EQ:It_simple}
\begin{split}
I_T = &\int^{+3}_{-3}\lambda(u_{y_1})\int_{0}^{+3}\hat{\lambda}(u_{x_1})\int_{u_{x_1}}^{+3}\left[\frac{1-e^{-\frac{u_{r_1}^2}{2}}}{u_{r_1}}\right]\frac{u_{x_1}}{u_{r_1}}u_{x_2}\hat{\lambda}(u_{x_2}) du_{x_2} du_{x_1} du_{y_1} \\
&- \int^{+3}_{-3}\lambda(u_{y_2})\int_{0}^{+3}\hat{\lambda}(u_{x_1})\int_{u_{x_1}}^{+3}\left[\frac{1-e^{-\frac{u_{r_2}^2}{2}}}{u_{r_2}}\right]\frac{u_{x_2}}{u_{r_2}}u_{x_1}\hat{\lambda}(u_{x_2}) du_{x_2} du_{x_1} du_{y_2}\\
&\approx0.0325
\end{split}
\end{equation}
The integral is independent of $\sigma_x$. And the $\mathcal{W}'_{sc}$ can be expressed in a simplified form as:
\begin{equation}
\label{EQ:w'sc}
   \mathcal{W}'_{sc}= 0.065\frac{N_er_e}{\sqrt{\pi}\gamma^3\sigma_z}.
\end{equation}
\begin{figure} 
\begin{center}
\includegraphics[height=6cm]{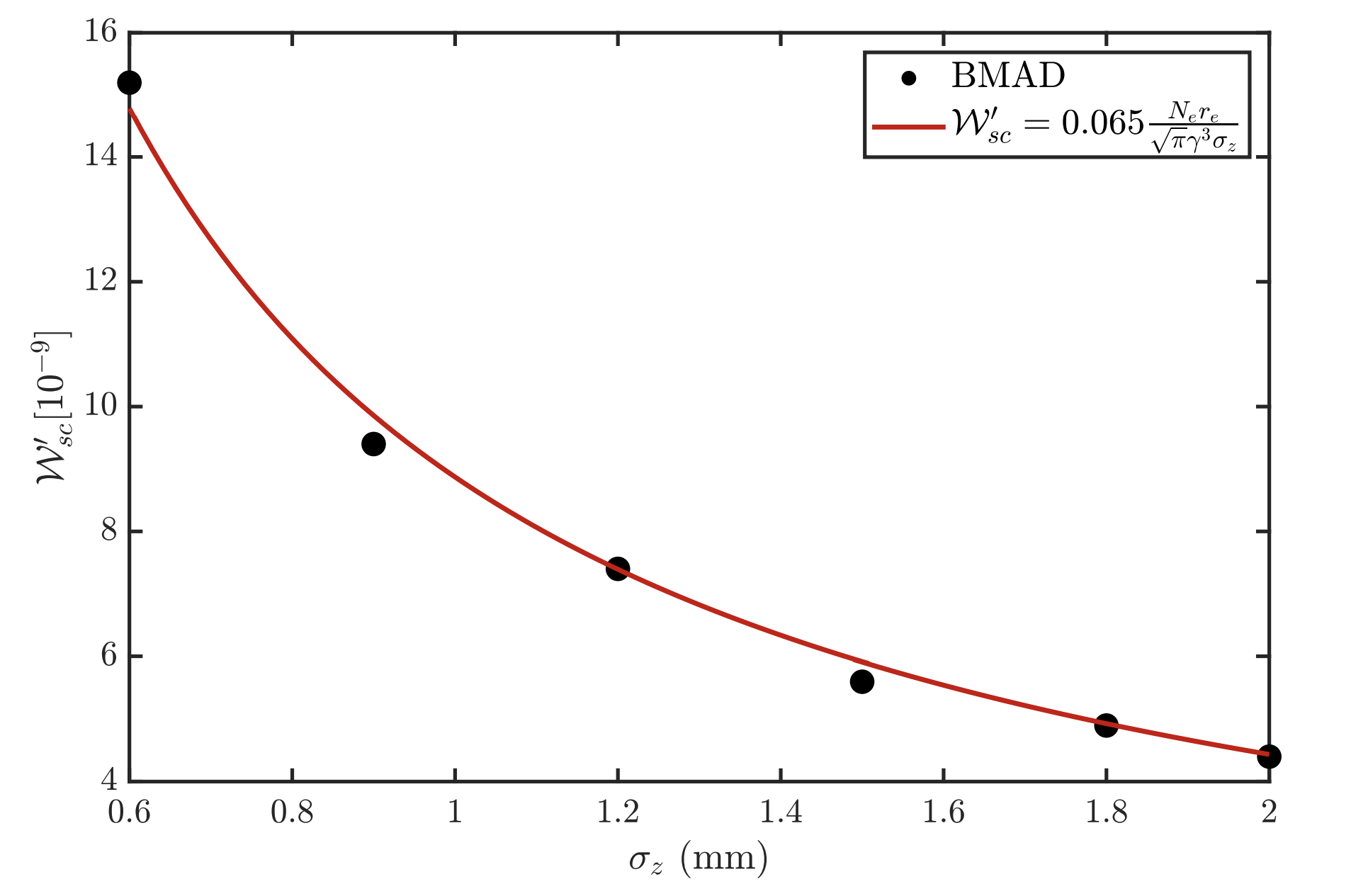} 
\caption{\label{fig:theory_mean_emit}
The analysis results of Eqs.(\ref{EQ:w'sc}) and the simulation results for the variance in mean emittance caused by TSC.}
\end{center}
\end{figure}
The simulation results from BMAD were used to verify the Eqs.(\ref{EQ:w'sc})\cite{Mayes,Sagan2006}. The bunch charge is 100pC, the bunch length is 0.6mm, and $\sigma_x=1\mathrm{mm}$. As shown in Fig.\ref{fig:theory_mean_emit}, there is good agreement between the analytical and simulation results for this typical case in ERL injectors. Eqs.(\ref{EQ:w'sc}) demonstrates that $\mathcal{W}'_{sc}$ is determined by three factors: the bunch length, the energy, and the bunch charge. This implies that for a round beam, the variance is uniform. In the case of a beamline without variance in energy and bunch length, the variance along the beamline is

\begin{equation}
\label{EQ:wsc}
   \mathcal{W}_{sc}= 0.065\frac{N_er_e\Delta l}{\sqrt{\pi}\gamma^3\sigma_z}.
\end{equation}
Where $\Delta l$ is the length of the beamline. Although the assumption of a round beam does not hold in the matching line and mergers, this estimate can still provide a valuable reference for the optimization processes of the injection section. The optimization target can be set as the inverse of Eqs.\ref{EQ:wsc} for pre-modulation, aiming to counteract the downstream non-linear distribution in transverse phase space caused by TSC. This inverse modulation is derived from the longitudinal gradient of the solenoid field (refer to Eqs.(14)-(16) in Ref\cite{huang2008analysis}).

\section{the displacements caused by LSC in the merger sections}\label{S3}

In machines with circular beamlines, a merger section is essential for transitioning the beam from the injection section to the circulator. ERL systems typically operate with injector exit energies below 20 MeV to optimize energy efficiency. As a result, the bunches in the merger operate within the space charge regime. Unlike the straight segments found in the injector or matching line, the merger operates as a deflection line characterized by non-zero dispersion. And the additional energy sprad will coupling with the dispersion lead to the displacements at the end of merger sections. These displacements further induce growth in projected emittance (a simple schematic illustration can be found in Fig.4 of Ref.\cite{hwang2012effects}). 

Relevant studies have been conducted to suppress this emittance growth\cite{litvinenko2006merger,kayran2005method,hajima2004emittance,hounsell2021optimization,hwang2012effects,kuske2010merger}. The zigzag merger and Triple Bend Achromat (TBA) merger have demonstrated superior performance in suppressing the emittance growth caused by LSC\cite{litvinenko2006merger, chen2024inj}. However, both types have their own limitations: For the zigzag merger, its compact structure limits its application in practical machines. For the TBA merger, the near-zero $R_{56}$ implies higher requirements for the injection sections if the target bunch length is shorter. On the other hand, a quantitative calculation of the displacements caused by LSC for any bunch is necessary. It can help us estimate the performance of mergers in the design work without relying on simulations. Optimization for individual bunches is time-consuming and lacks universality. The quantitative calculation would significantly reduce the time required for optimization. The R-matrix expansion to $s^3$ method was proposed for quantitative calculation of LSC effects\cite{hwang2012effects}. However, the coefficients of this method need to vary significantly for different bunches. The central symmetry in the transfer matrix can qualitatively describe the performance of the merger section, and the zigzag merger was proposed based on this method\cite{kayran2005method,litvinenko2006merger}. However, this approach only considers the additional energy spread caused by LSC to the second order. As discussed in the following section of this study, the performance of the zigzag merger deteriorates when the initial energy chirp cannot be neglected. Similarly, qualitative theories based on the integral method of Coherent-Synchrotron Radiation (CSR) are only applicable when the integral value is zero\cite{chen2024inj,venturini2016design}. It's important to note that the simulations in this section, which were conducted to compare with theoretical predictions, only considered the SC effects in drift sections. And the order of transfer matrix in simulation is set to 1.

In this section, the integral method in Ref\cite{chen2024inj} is further developed for quantitative calculation of displacements caused by LSC. The simulation results from BMAD show good agreement with theoretical predictions. Additionally, the merger based on the previous design has been optimized using this method. Compared to the zigzag merger, it demonstrates superior performance when the initial bunch characteristics cannot be neglected.

Similar with the CSR in the bend magnets, the displacements caused by LSC at the exit of beamline without dispersion can be expressed as
\begin{equation}
    \begin{aligned}
& \hat{x}_{L S C}(u)=\int_{s_0}^{s_f} \delta_{L S C}^{\prime}(u,s) \cdot R_{16}^{s \rightarrow s_f} d s \\
& \hat{x}_{L S C}^{\prime}(u)=\int_{s_0}^{s_f} \delta_{L S C}^{\prime}(u,s) \cdot R_{26}^{s \rightarrow s_f} d s.
\end{aligned}
\label{EQ:disp_LSC}
\end{equation}
$R_{ij}^{s\rightarrow s_f}$ is the transfer matrix from $s$ to $s_f$, where $s_f$ is the exit of the beamline. $\delta_{L S C}^{\prime}(u,s)$ is the additional energy spread caused by LSC at point $s$. Furthermore, the bunch length is variable during mergers in most cases. The longitudinal position $z$ in the bunch coordinate coordinate in Ref\cite{chen2024inj} is replaced by the slice index $u=\frac{z}{\sigma_z}$. This index is invariant under Lorentz transformations. The integral in bends can be neglected when the merger satisfies the condition $\int_{B} R_{5i} ds \ll \int_{all} R_{5i} ds$, where $\int_B$ represents the integral along the bends in the merger, and $\int_{all}$ denotes the integral over the entire merger region. This approximation can greatly simplify the calculation process because the $R_{i6}^{s\rightarrow s_f}$ in the non-bend region is a constant. And the Eqs.(\ref{EQ:disp_LSC}) can be replaced by

\begin{equation}
    \begin{aligned}
\hat{x}_{_{LSC}}(u) &=\sum_i^n R_{16,i}^{s\rightarrow s_f}\int\delta'(u)ds=\sum_i^n R_{16,i}^{s\rightarrow s_f}\Delta \delta_{_{LSC},i}(u) \\
\hat{x}_{_{LSC}}^{\prime}(u) &=\sum_i^n R_{26,i}^{s\rightarrow s_f}\int\delta'(u)ds=\sum_i^n R_{16,i}^{s\rightarrow s_f}\Delta \delta_{_{LSC},i}(u).
\end{aligned}
\label{EQ:disp_LSC2}
\end{equation}
The index $i$ represents the sequence number of the bends in mergers, and the $n$ is the total number of bends in merger. $\delta_{_{LSC},i}(u)$ is the variance in energy spread between the $i$-th and the $(i+1)$-th bend. In the design work, we can use the symmetric condition to:

\begin{equation}
\label{EQ:symm}
    R_{16}^{s\rightarrow s_f}=-R_{52}^{s_f\rightarrow s}\\
    R_{26}^{s\rightarrow s_f}=-R_{51}^{s_f\rightarrow s}.
\end{equation}

Where $R_{51}^{s_f\rightarrow s}$ and $R_{52}^{s_f\rightarrow s}$ simplified to $R_{51}$ and $R_{52}$ in this paper. It's important to note that the optimization process, after substitution, proceeds in reverse order. This means the sequence of elements in the optimization is inverted compared to their actual arrangement. For example, the first bend encountered in the optimization corresponds to the last bend in the physical setup. The reverse merger process is referred to as "re-merger" in this study.

In the paraxial approximation, the additional energy spread caused by LSC in drift is

\begin{equation}
    \delta'_{_{LSC}}(u)=\frac{eE_z(u)}{\gamma m_ec^2}.
 \end{equation}

For any distribution of the bunch, the $E_z(u)$ is
\begin{equation}
    \begin{split}
        E_z(u) =& -\frac{1}{\gamma^2\sigma_z^2}\frac{\partial}{\partial u}\int_{-\infty}^{\infty}\frac{\rho(u_x',u_y', u')}{\sqrt{(u_x-u_x')^2+(u_y-u_y')^2+(u-u')^2}}\,du_x'\,du_y'\,du' \\
        =& \frac{1}{\gamma^2\sigma_z^2}\int_{-\infty}^{\infty}\rho(u_x',u_y', u')\cdot\frac{(u-u')}{[(u_x-u_x')^2+(u_y-u_y')^2+(u-u')^2]^{3/2}}\,du_x'\,du_y'\,du',
    \end{split}
\end{equation}
where $u_x=\frac{x}{\sigma_x}\frac{\sigma_x}{\gamma \sigma_z}$ and similar formula for $u_y$. For the long bunch approximation, we have the following relation for any slice index $u$:
\begin{equation}
\label{EQ:approxEz}
   \frac{ \delta'_{_{LSC}}(\sigma_{z_1})}{\delta'_{_{LSC}}(\sigma_{z_2})}\approx \frac{\sigma_{z_2}^2}{\sigma_{z_1}^2}.
\end{equation}
Fig.(5) shows the normalized variance in energy spread from BMAD simulations. The relationship between the variance and the bunch length approximately follows the relation in Eqs.(\ref{EQ:approxEz}). The bunch charge is 100pC, the initial normalized transverse emittance is $0.5\mathrm{\mu mrad}$ and the initial $
\beta_{x\&y}$ with the Range in [10, 100]. These are typical parameters for ERL mergers.

\begin{figure} 
\begin{center}
\includegraphics[height=6cm]{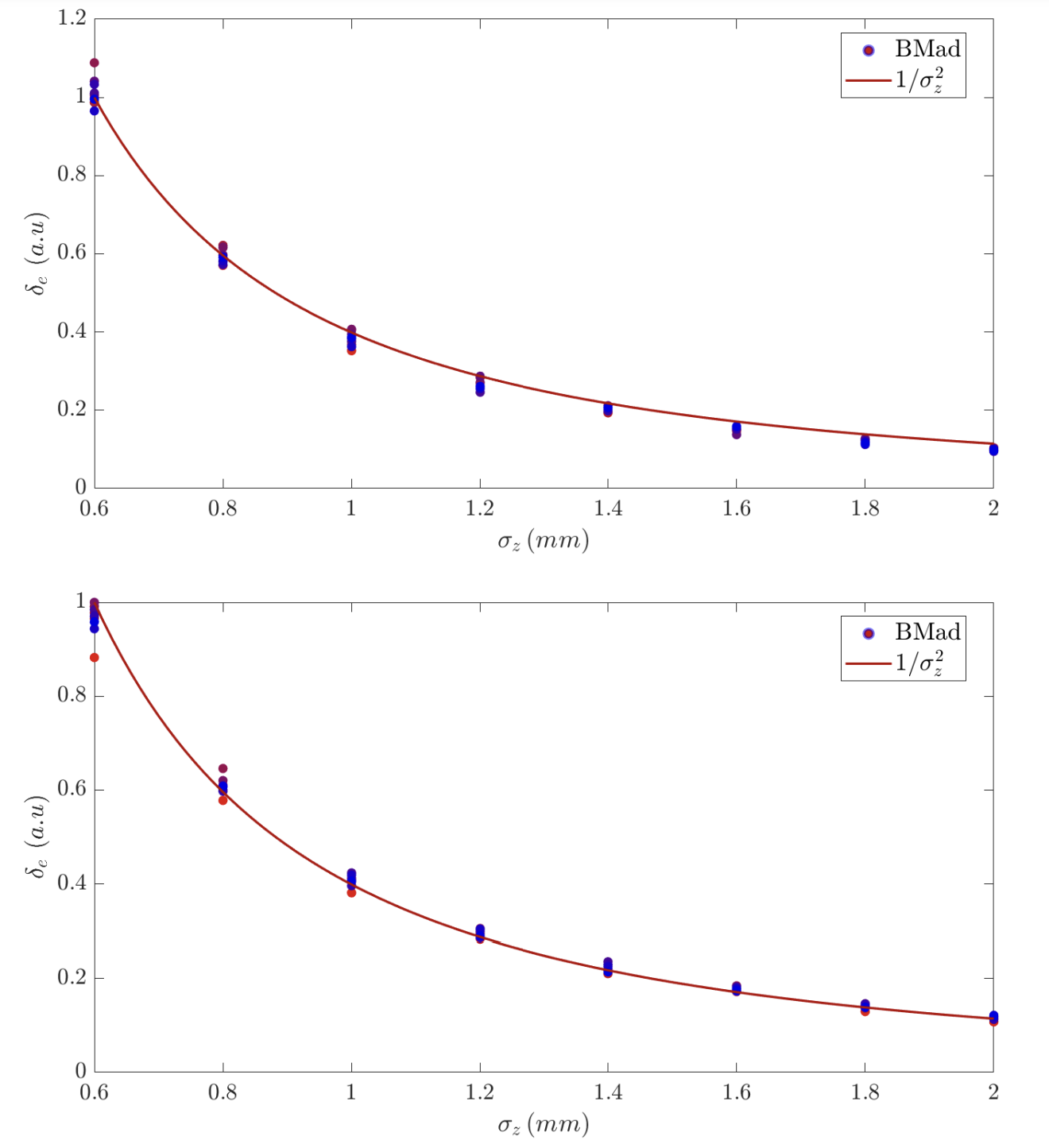} 
\caption{\label{fig:delta_vs_sigz}
The relationship between the variance in energy spread caused by LSC and the bunch length in Lab frame, (top):10MeV, (bottom): 5MeV }
\end{center}
\end{figure}

\begin{figure} 
\begin{center}
\includegraphics[height=6cm]{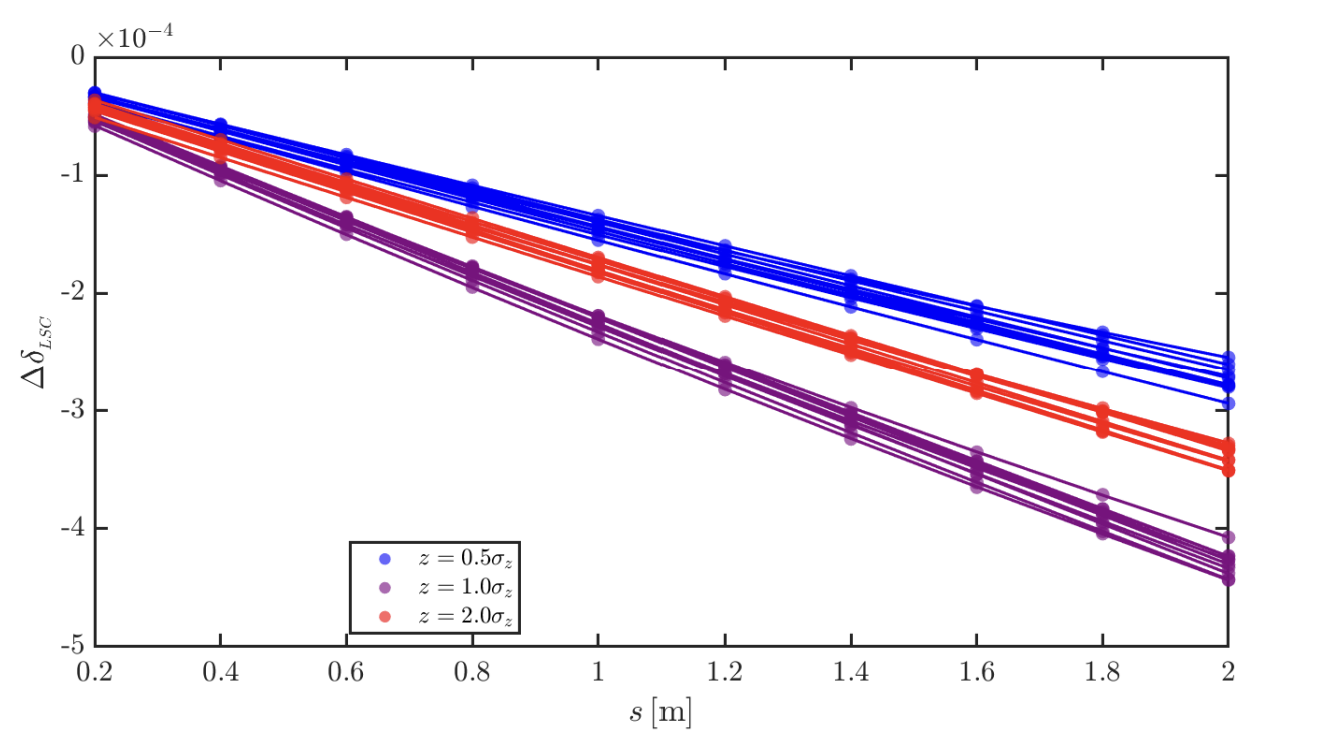}
\caption{\label{fig:evo_delta}
The evolution of $\Delta \delta_{_{LSC},i}$ for different slice in BMAD and 2nd order fit curve. }
\end{center}
\end{figure}

With the approximate relation in Eq.(\ref{EQ:approxEz}), the $\Delta \delta_{_{LSC}}$ in Eqs.(\ref{EQ:disp_LSC2}) in the non-bend region can be expressed as
\begin{equation}
    \label{EQ: DELTA_LSC}
    \Delta \delta_{_{LSC},i}(u)=\frac{k_{sc}(u)}{\sigma_{z}^2}\sum_{j=1}^n a_j(u)s^j.
\end{equation}
As the high order R-matrix method in Ref\cite{hwang2012effects}, the  $\Delta \delta_{_{LSC},i}$ can be expressed as a series expansion. And the variance in bunch length in non-bend region is much smaller than the variance in bends, so the bunch length in the non-bend region is a constant. $k_{sc}$ is determined by the bunch distribution. The $a_j(u)$ are the coefficients of the j-th terms. In this study, the matching line between bends in the merger section is less than 2m. The $\Delta \delta_{LSC}$ with a 2m drift for different slices is shown in Fig.2. To evaluate the effects with different evolutions of beam size, the initial $\alpha_x$ is set within the range of [-10, 10].$ \alpha_y$ is set to be the opposite of $\alpha_x$. As shown in Fig. 5, the evolution for different cases tends to be linear. The maximum second-order term is only $5\%$ of the first-order term in the fit curve. So the $\delta'_{_LSC}(u)$ is a constant in these cases. 

When considering only linear compression, the bunch length at any point $s$ in the re-merger can be expressed as:
\begin{equation}
    \label{EQ:SIGZ}
\sigma_z(s)=\sigma_{z_0}\left[1+h(R_{56,tot}-R_{56}(s))\right],
\end{equation}
where 
\begin{equation}
    h=\frac{\sigma_e}{\sigma_{z_0}}
\end{equation}
is the initial energy spread. $\sigma_e$ is the initial energy spread. $\sigma_{z_0}$ is the initial bunch length. $R_{56}(s)$ and $R_{56,tot}$ are the $R_{56}$ at point $s$ of re-merger and the total $R_{56}$ of the merger. Combining Eqs.(\ref{EQ:disp_LSC2}), (\ref{EQ:symm}), and (\ref{EQ:SIGZ}), the displacement caused by longitudinal space charge (LSC) at the exit of the merger can be expressed as

\begin{equation}
    \label{EQ:dis_LSC}
    \hat{x}_{_{LSC}}(u)=-\frac{k_{sc}(u)}{\sigma_{z_0}^2}\sum_{i=1}^{n}\frac{R_{52,i}\cdot L_i}{[1+h(R_{56,tot}-R_{56,i}]^2},
\end{equation}

\begin{figure} 
\begin{center}
\includegraphics[height=12cm]{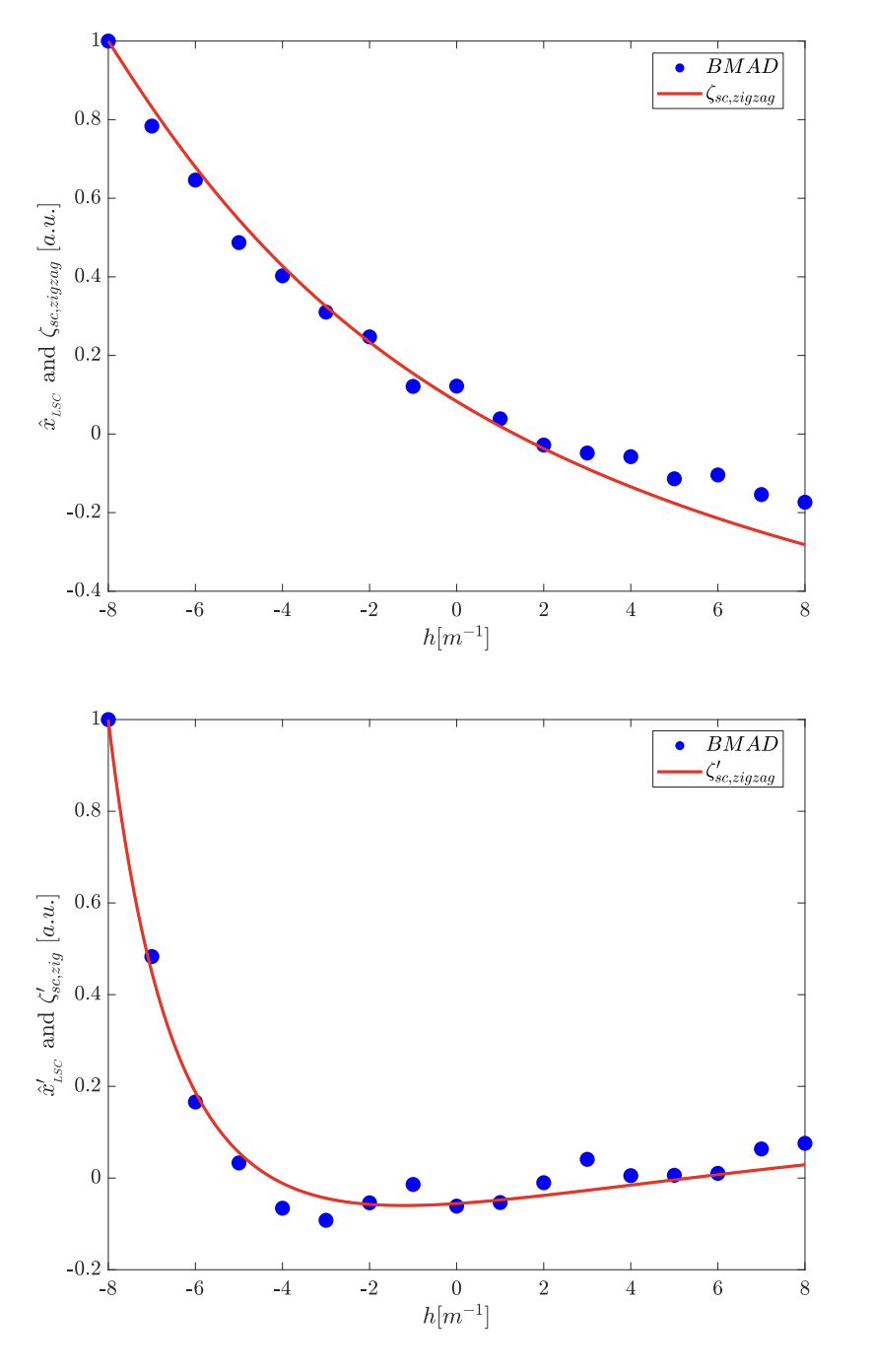}
\caption{\label{fig:zig_disp}
The normalized displacements in $x$-axis(top) and the $x'$-axis(bottom) caused by LSC. The bunch charge is 100pC, the initial bunch length is 1mm, the energy is 10MeV and the initial emittance is 0.5$\mathrm{\mu mrad}$.}
\end{center}
\end{figure}

\begin{equation}
    \label{EQ:disp_LSC3}
    \hat{x}'_{_{LSC}}(u)=-\frac{k_{sc}(u)}{\sigma_{z_0}^2}\sum_{i=1}^{n}\frac{R_{51,i}\cdot L_i}{[1+h(R_{56,tot}-R_{56,i}]^2}.
\end{equation}
Where $R_{5j,i}$ is the $R_{5j}$ after the $i$-th of re-merger, $L_i$ is the length between the $i$-th and $(i+1)$-th bend of re-merger.As discussed in Eqs. (\ref{EQ:dis_LSC}) and (\ref{EQ:disp_LSC3}), the displacements in $x$ and $x'$ can be decomposed into two independent components: the coefficient $\frac{k_{sc}(u)}{\sigma_{z_0}^2}$, which is solely determined by the parameters of the initial distribution, and the summation term, which is independent of the distribution characteristics. Interestingly, it should be noted that the parameter $h$ in these equations is also independent of the distribution. Then we defined the summation term as modified $\zeta_{sc}$ functions (simplified as $\zeta_{sc}$ function in this study) based on the $\zeta_{sc}$ function in Ref\cite{chen2024inj}: 
\begin{equation}
    \label{eq:zeta}
    \zeta_{sc}=\sum_{i=1}^{n}\frac{R_{52,i}\cdot L_i}{[1+h(R_{56,tot}-R_{56,i}]^2},
\end{equation}

\begin{equation}
    \label{eq:zeta'}
    \zeta'_{sc}=\sum_{i=1}^{n}\frac{R_{51,i}\cdot L_i}{[1+h(R_{56,tot}-R_{56,i}]^2}
\end{equation}
A smaller $\zeta_{sc}$ function leads to smaller displacements for any bunches. Furthermore, the $\zeta_{sc}$ function varies for different values of $h$, unless the $R_{56}$ of the merger is close to zero, as in the case of the TBA merger in Ref\cite{chen2024inj}. For the zigzag merger, the $\zeta_{sc}$ close to zero when $h=0$. And the parameters for $\zeta_{sc}$ are
\begin{equation}
    \begin{gathered}
R_{51,1}=-\sin \theta_1, \quad R_{52,1}=-R(1-\cos \theta_1) \\
R_{51,2}=\sin \theta_1 \\
R_{52,2}=\frac{1}{2} \sec \theta_1^2 \sin \frac{\theta_1}{2}\left(8 d_1 \cos \frac{\theta_1}{2}+R\left(-2 \sin \frac{\theta_1}{2}+5 \sin \frac{3 \theta_1}{2}+3 \sin \frac{5 \theta_1}{2}\right)\right) \\
R_{51,3}=-\sin \theta_1 \\
R_{52,3}=\frac{1}{2} \sec \theta_1^2 \sin \frac{\theta_1}{2}\left(16 d_1 \cos \frac{\theta_1}{2}+R\left(-11 \sin \frac{\theta_1}{2}+5 \sin \frac{3 \theta_1}{2}+3 \sin \frac{5 \theta_1}{2}\right)\right).
\end{gathered}
\end{equation}
$\theta_1=\frac{\theta_2}{2}=10^{\circ}$ in the most cases, the $d_1=\frac{d_2}{2}=0.6$m in this study and the radius of the bends $R$ are set to 0.5m. The normalized displacements from BMAD simulation and Eqs.(\ref{eq:zeta}) and (\ref{eq:zeta'}) for zigzag merger is shown in Fig.\ref{fig:zig_disp}. The slice index $u \approx 1.1$ corresponds to the maximum value of the LSC force.  

For cases where $h > 0$, the bunch length increases after passing through the zigzag mergers, while it decreases for $h < 0$. As illustrated in Fig.\ref{fig:zig_disp}, the zigzag merger demonstrates superior performance in scenarios with minimal bunch compression or in decompression cases. For cases where $h < -4$, the performance in suppressing LSC effects begins to deteriorate, with this degradation being particularly noticeable along the $x'$-axis. Reducing the length of $d_1$ can significantly enhance the performance of the zigzag structure\cite{chen2024inj}; however, this results in a more compact configuration. Then we maintain the last bending magnet of the merger (which is also the first bend of the re-merger) and the length of the upstream drift as constants, while adjusting the subsequent structures. A schematic diagram of the new merger design is illustrated in Fig.\ref{fig:newmer}. To achieve achromatic, the $l_2$ meets:

\begin{figure} 
\begin{center}
\includegraphics[height=6cm]{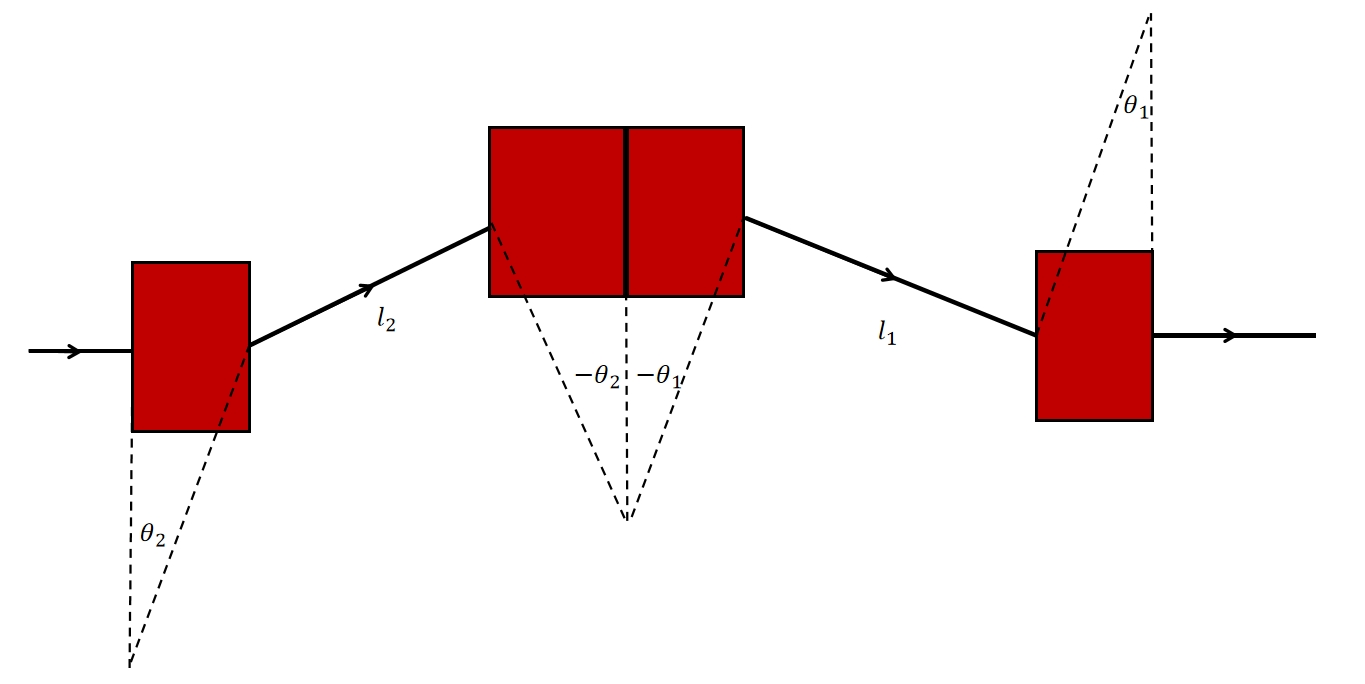}
\caption{\label{fig:newmer}
The schematic diagram of the new merger design. $\theta_1=10^{\circ}$ and $l_1=0.6$m.}
\end{center}
\end{figure}

\begin{figure} 
\begin{center}
\includegraphics[height=12cm]{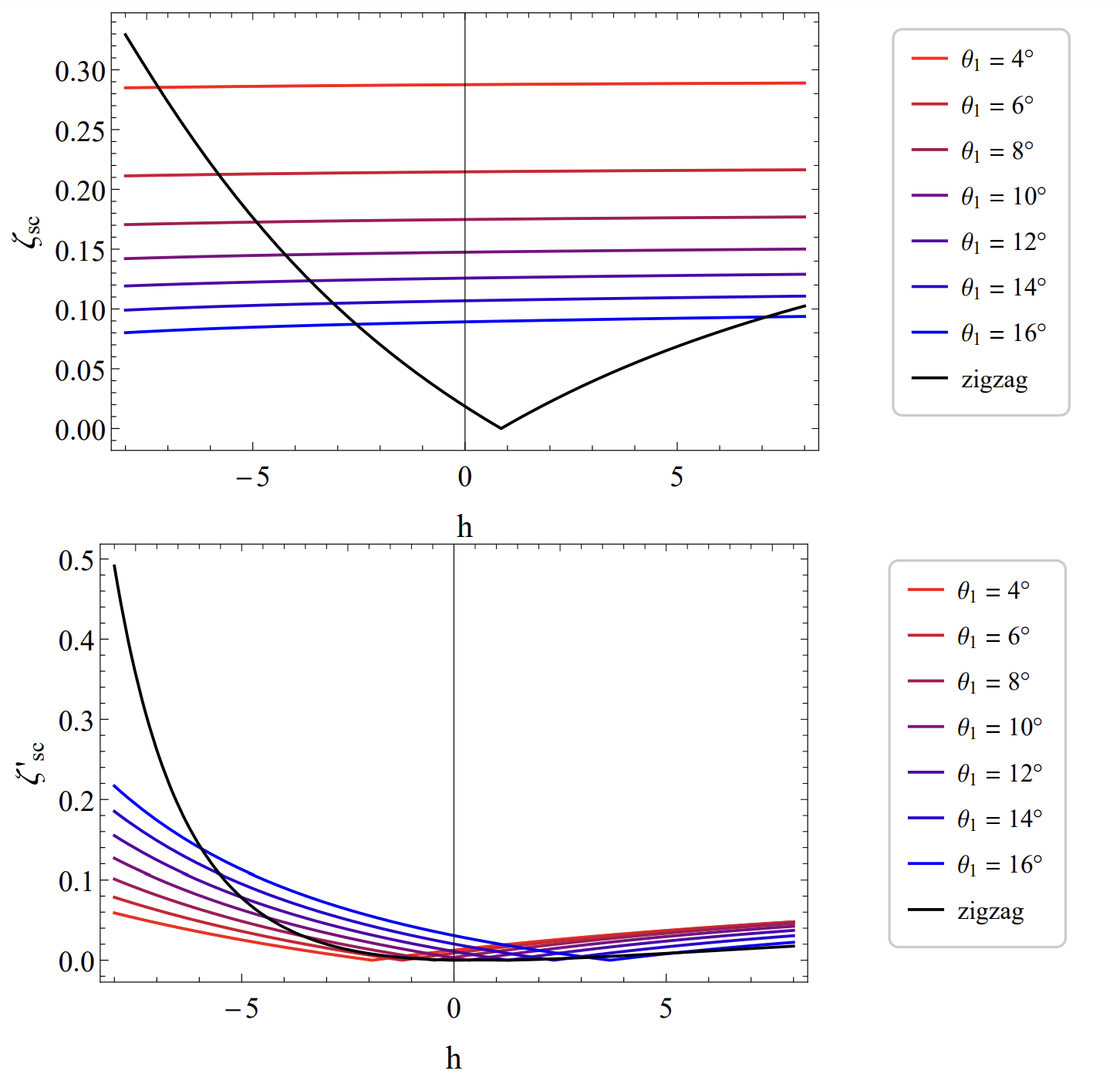}
\caption{\label{fig:newmerger_zeta}
The $|\zeta_{sc}|$ functions of the new merger with different $\theta_2$. The black curves are the $|\zeta_{sc}|$ function of zigzag merger.}
\end{center}
\end{figure}

\begin{figure} 
\begin{center}
\includegraphics[height=12cm]{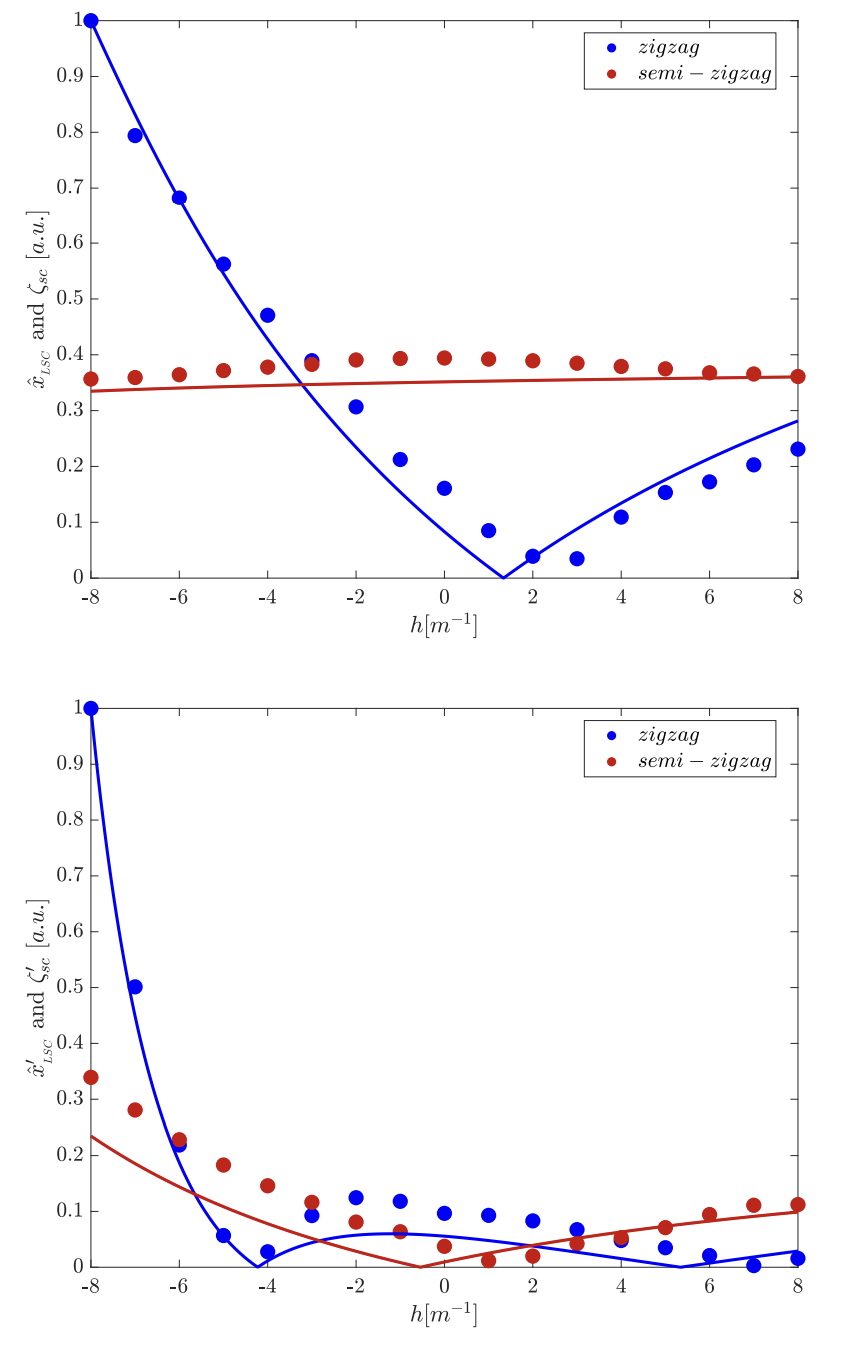}
\caption{\label{fig:newmervszig}
The normalized displacements in $x$-axis(top) and the $x'$-axis(bottom) caused by LSC for zigzag and semi-zigzag merger. The bunch parameters are similar as the bunch in Fig.\ref{fig:zig_disp}}
\end{center}
\end{figure}

\begin{equation}
    \label{EQ:l3}
    l_2=\operatorname{cot}\theta_ 2(-2 R+\operatorname{cos}\theta_ 2\operatorname{sec}\theta_1(2 R+l_1 \operatorname{tan}\theta_1)).
\end{equation}
Notice the radius of the bends in merger is a constant. And the $\zeta_{sc}$ functions of the new merger with different $\theta_2$ are shown in Fig.\ref{fig:newmerger_zeta}. From the figure, it can be seen that when $h>-3$, the zigzag merger shows superior performance. Especially for the displacements in $x$-axis. However, as $h$ decreases below -4, the performance of the zigzag merger deteriorates. The new merger maintains better performance, thus demonstrating a superior ability in suppressing LSC in these cases. Furthermore, the performance of these mergers varies significantly with changes in $\theta_2$. In cases with larger $\theta_2$ values, the performance improves in $x$ direction and worsens along the $x'$ direction. Designers need to make trade-offs based on the specific cases.

In this study, $\theta_2 = \theta_1$. The new merger is a chicane with a small angle, and the spacing between the middle bends is zero. On the other hand, this new merger is exactly half of a zigzag merger, hence it is called a semi-zigzag merger. And the simulation results for zigzag and semi-zigzag mergers is shown in Figs\ref{fig:newmervszig}.  Compared to the zigzag merger, the semi-zigzag merger requires a smaller deflection angle for the high-energy beam. For a zigzag merger without intersection in the beamline, the high-energy beam needs to be deflected by $10^{\circ}$. This is a challenging task for high-energy machines. Furthermore, cases where $h<-4$ (which corresponds to a bunch with 1mm length and 0.004 in energy spread) are common. The semi-zigzag merger is more suitable for these situations.

\section{conclusion and discussion}

In this study, two methods were optimized to address SC effects downstream of ERL injectors. For the TSCF, the mean emittance was specifically calculated for 3-D Gaussian bunches, which is the common situation at the exit of injection sections. The developed mean emittance helps designers estimate the TSC effect in downstream sections in advance. By pre-modulating the bunches at the end of the injection section, growth in slice emittance in downstream parts can be suppressed. For the LSC, the $\zeta_{sc}$ was further developed, enabling quantitative analysis of LSC in mergers. Based on this method, the semi-zigzag merger was proposed. For bunches with larger initial energy chirp, it demonstrates better performance in suppressing the transverse displacements caused by LSC. Additionally, this method is applicable to any merger and can be used to estimate the effects caused by LSC for other mergers, simplifying the design process. An S2E injector simulation based on these two methods is currently in progress. At the same time, more precise calculations for pre-modulation, such as further optimization of solenoids, need to be supplemented in future work.

\begin{acknowledgments}

\end{acknowledgments}
This work was supported by the CAS project for Young Scientists in Basic Research (YSBR-042), The National Natural Science Foundation of China (12125508, 11935020), Program of Shanghai Academic Technology Research Leader (21XD1404100), and Shanghai Pilot Program for Basic Research Chinese Academy of Sciences, Shanghai Branch (JCYJ-SHFY-2021-010), the Natural Science Foundation of Shanghai (22ZR1470200).


\providecommand{\noopsort}[1]{}\providecommand{\singleletter}[1]{#1}%

\end{document}